\documentstyle[aps,twocolumn,prl,epsf]{revtex}

\begin{document}

\newcommand{\be}{\begin{equation}} 
\newcommand{\ee}{\end{equation}}
\newcommand{\bn}{\begin{eqnarray}}
\newcommand{\en}{\end{eqnarray}}
\newcommand{\bq}{\begin{eqnarray}}
\newcommand{\eq}{\end{eqnarray}}

\draft

\twocolumn[\hsize\textwidth\columnwidth\hsize\csname @twocolumnfalse\endcsname

\draft

\title{On the heavy-fermion behavior of the pyrochlore transition-metal oxide 
 $LiV_{2}O_{4}$}

\author{M. S. Laad$^1$, L. Craco$^2$ and E. M\"uller-Hartmann$^1$}
\address{${}^1$Institut f\"ur Theoretische Physik, Universit\"at zu K\"oln,
 77 Z\"ulpicher Strasse, D-50937 K\"oln, Germany \\
${}^2$ Max Planck Institute for the Physics of Complex Systems,
D-01187 Dresden, Germany
} 

\date\today

\maketitle

\begin{abstract}
Motivated by the heavy fermion Fermi liquid (HFFL) features observed at 
low-$T$ in the pyrochlore $LiV_{2}O_{4}$, we consider a material-specific 
model that includes aspects of the local quantum chemistry, the geometrically 
frustrated lattice structure, and strong correlations in a {\it single} 
approach.  In particular, we show how geometrical frustration (GF) gives 
rise to a crossover scale, $T^{*}<<J$, the intersite (AF) exchange, below 
which the metallic system shows HFFL features. Our scenario is a specific 
realization of the importance of GF effects in driving HFFL behavior in 
$LiV_{2}O_{4}$, and provides a natural understanding of 
various puzzling features observed experimentally. 
\end{abstract}
     

]

 Systems of the type $AB_{2}O_{4}$ with a magnetic $B$ site 
consistently show strange properties. In particular, systems with 
$B=Ti,V,Cr$ are interesting because the $d$ electrons occupy $t_{2g}$ 
orbitals that do {\it not} hybridize with the oxygen 
orbitals along certain directions\cite{[1],[2],[3],[4],[5]}. In the geometrically frustrated 
structure, the nearest neighbor exchange interaction dominates with a 
maximization of frustration.  Concomitantly, only $3d$ bands ($t_{2g}$) 
cross the Fermi level $(\mu)$, leading one to 
have to deal with the ubiquitous strong correlations. With a divalent 
$A$-ion, the $B$ site has integral valence, giving rise to Mott insulating 
behavior with frustrated magnetism, e.~g., in $ZnV_{2}O_{4}$. On the other 
hand, with a monovalent $A$ site, $B$ is mixed-valent, resulting in a 
narrow-band metal.  This combination of frustrated magnetism and correlated 
metallic behavior leads to a range of complex and unusual manifestations, 
from superconductivity, for example in $LiTi_{2}O_{4}$, via heavy-fermion 
Fermi liquid (HFFL) behavior in $LiV_{2}O_{4}$, to spin-glass behavior in 
other systems.

The best example in the second category above is 
$LiV_{2}O_{4}$~\cite{[6]}, a paramagnetic metal for $T > 0.01$K. 
The low-$T$ Sommerfeld constant achieves its highest value for a 
$d$-electron system, $0.42 J/mole/K^{2}$ at $T=1.5K$. The resistivity, 
the Woods-Saxon ratio, as well 
as the Wilson ratio all exhibit behaviors expected for prototypical 
rare-earth based HFFL metals~\cite{[6]}.
Importance of geometrical frustration (GF) for magnetic properties is 
shown by the absence of a low-$T$ magnetic instability, and  
further vindicated by neutron scattering (INS) studies~\cite{[7]}, which reveal a 
response characteristic of {\it insulating}, frustrated magnets.  
This leads one to ask:  What is the role of GF in
driving HFFL behavior?  More precisely, how does one reconcile the 
HFFL behavior in thermodynamics and transport with a magnetic
response characteristic of frustrated, {\it insulating} magnets?
What is the role of strong correlations in the $3d$ bands?  How do these two
subsystems couple to (affect) each other at low-$T$?

Much attention has been devoted to these issues. LDA bandstructure 
calculations indeed show that the $3d$ $t_{2g}$ bands alone cross the Fermi
level~\cite{[8]}. A trigonal distortion splits the three-fold degenerate 
$t_{2g}$ states into the (lower lying) singlet $A_{1g}$ with a bandwidth of 
$1eV$, and an $E_{g}$ doublet with a bandwidth of $2eV$ in the solid.  Given 
the formal $d^{1.5}$ state of $V$, the $A_{1g}$ band is half-filled, and the 
$E_{g}$ bands are quarter-filled, leading~\cite{[9]} to suggestions that 
an effective Anderson model could be used. However, approaches  
along these lines require the rather ad-hoc introduction of a large 
intersite Kondo coupling to counterbalance the strong local Hund's rule 
coupling, the origin of which is unclear.  Moreover, frustration effects 
are not important in these pictures.  On the other side, Fulde 
{\it et al.}~\cite{[10]} have proposed that the $V$ lattice of corner-sharing 
tetrahedra frustrates charge ordering and leads instead to isolated finite 
chains of $S=1/2$ and $S=1$.  The gapless fermionic spin excitations 
of the $S=1/2$ chains give the large $\gamma$ coefficient of the low-$T$ 
specific heat.  This picture needs to be extended to derive the HFFL 
properties, and to make a detailed comparison with INS results. 
Varma~\cite{[11]} seeks to understand the HFFL properties by analyzing the 
energetics of the crossover which a $S \ne 1/2$ impurity must undergo in 
the Kondo effect. Frustration does not play any role here either. Recently, 
Burdin {\it et al.}~\cite{[12]} have studied the role of frustration effects 
in driving HFFL behavior. However, frustration effects are put in by hand, 
limiting a direct comparison with the actual material.  Lacroix~\cite{[12b]} has recently
sketched the outlines of a picture for HFFL behavior in $d$-band oxides with
geometrical frustration.  

  Here, we study a theoretical model that is material-specific and
 includes {\it all} the 
relevant degrees of freedom and explicitly address the questions
above.  In particular, we show how the HFFL properties are reconciled with
the magnetic response characteristic of frustrated, insulating magnets. 
The choice of the model Hamiltonian is motivated by experimental 
constraints~\cite{[6]} as well as by results of first-principles LDA 
calculations~\cite{[8]}: 

\be
H=H_{s}+H_{b}+H_{sb}\;,
\ee
where $H_{s}=\sum_{ij}J_{ij}{\bf S}_{i}.{\bf S}_{j}$ is the 
Heisenberg-like $S=1/2$ 
Hamiltonian describing the localized spin degrees of freedom originating 
from the half-filled narrow singlet $A_{1g}$ band.  In $LiV_{2}O_{4}$, the 
next-nearest neighbor coupling may be important, though its magnitude is 
not reliably known.

\bn
\nonumber
H_{b} &= & -\sum_{<ij>}t_{ab}(c_{ia\sigma}^{\dag}c_{jb\sigma}+h.c) 
+ U\sum_{ia}n_{ia\uparrow}n_{ia\downarrow} \\
& + & U_{ab}\sum_{i,a,b}n_{ia}n_{ib}
\en
describes the electronic degrees of freedom for the $E_{g}$ bands.  
And 
$H_{sb}=-J_{H}\sum_{i,a,b}{\bf S}_{i}.({\bf \sigma}_{ia}
+{\bf \sigma}_{ib})$ 
describes the coupling of these two subsystems via a local Hund's rule
interaction.

In what follows, $H$ is defined on the fully-frustrated 
pyrochlore (FFL) lattice,
entailing explicit consideration of geometrical frustration effects.
To proceed in this complicated situation, and motivated by 
observations, we make the plausible assumption: The underlying $A_{1g}$ 
spin configuration does affect the charge dynamics in the $E_{g}$ band, but 
that there is no qualitative change in the localized magnetic response coming
from the feedback effects due to this carrier dynamics.  This assumption is 
justified later.  With this, we follow the following strategy:
$(i)$ Treat the spin correlations in the Heisenberg like model on the 
FFL within a cluster approach capable of correctly treating the short-range
fluctuations which drive the spin liquid behavior as observed in INS 
experiments.
$(ii)$ Use the fact that the $A_{1g}$ spins are coupled to the correlated 
electrons in the $E_{g}$ band(s), and modify the hopping, $t_{ab}$, to 
solve the modified electronic model within the $d=\infty$ approximation, 
which is the best technique to reliably access dynamical effects of strong, 
local correlations~\cite{[14]}. As with the $A_{1g}$ spins, the $E_{g}$ 
electrons live on the  FFL, a fact that is indeed important for a consistent 
understanding, as we show below.  

  On the FFL, the electronic dispersion relation (or the spin-wave
dispersion for localized spins) has four branches corresponding to
the four-sublattice structure inherent in this geometry; two branches are
completely flat over the whole Brillouin zone, while the other two form 
dispersive bands in the solid.  Mathematically, 
$\epsilon_{\alpha}({\bf k})=2t$ for $\alpha=1,2$ and,
$\epsilon_{\alpha}({\bf k})=-2t (1 \pm 
\sqrt{c_{x}c_{y}+c_{y}c_{z}+c_{z}c_{x}}) $
for $\alpha=3,4$ $[c_{a} \equiv cos(k_{a}/2)]$.
Notice that the free dispersion alone gives a two-band structure in the 
unperturbed DOS (see Fig.~\ref{fig2}, with $U_{ab}=0$).  
For the $A_{1g}$ spins, one replaces $\epsilon({\bf k})$ by $J({\bf q})$ and 
$t$ by $J$ in the above equations. From the susceptibility data~\cite{[6]}, 
a large $g=2.23$ is inferred, suggesting strong {\it ferromagnetic} coupling 
between the carriers in the $E_{g}$ band and the ``localized'' spins in the 
$A_{1g}$ band (large $J_{H}$). So the carrier hopping rate is strongly 
coupled to, and reflects the underlying ($A_{1g}$) spin correlations: 
$t_{ab}({\bf S})\rightarrow t_{ab} 
\sqrt{1+<{\bf S}_{i}.{\bf S}_{j}>/2S^{2}}$~\cite{[17]}.

\begin{figure}[h]
\epsfxsize=3.4in
\epsffile{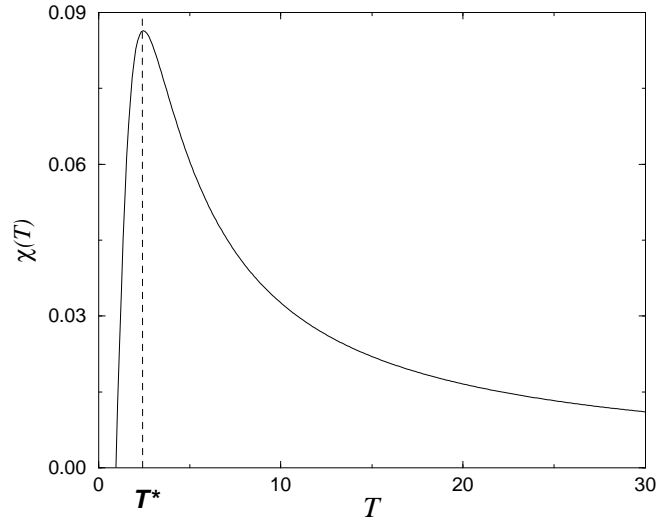}
\caption{Magnetic susceptibility of the localized spins in the $A_{1g}$
band of the pyrochlore lattice obtained from a self-consistent embedded
cluster approach~$[14]$. $T^\star$ marks the crossover 
temperature below which the system exhibits HFFL behavior.}
\label{fig1}
\end{figure}

\begin{figure}[h]
\epsfxsize=3.4in
\epsffile{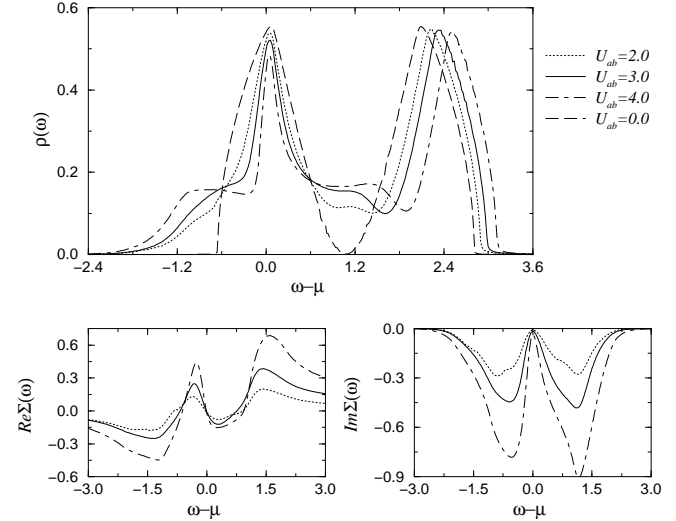}
\caption{Local spectral density (a), and real (b) and imaginary (c) 
parts of the s.p. self-energy for the Hubbard model on the 
fully-frustated pyrochlore lattice for diffetent values of the Coulomb 
interaction in the $E_g$ sector.}  
\label{fig2}
\end{figure}

  As mentioned before, we first focus on the $A_{1g}$ 
correlations, assuming, in accordance with LDA calculations, that this 
narrow band is completely occupied by one localized electron per site.  
Following~\cite{[15]}, we employ a self-consistently embedded cluster 
approach that treats the spin correlations in one tetrahedron {\it exactly} 
and mimics the influence of the remaining tetrahedra by inhomogeneous, 
selfconsistently determined magnetic fields. The resulting 
static spin susceptibility is shown in Fig.~\ref{fig1}.
$\chi_{s}(T)$ shows a Curie-Weiss (CW)-like form at 
high-$T$, followed by a maximum at lower $T^{*}<<J$.  And $\chi({\bf q},T)=g({\bf q})\chi(T)$, consistent with~\cite{[16]}.  At very low $T$, a 
gap opens up in the spin excitation spectrum; by fitting $\chi({\bf q},T)$ to  
an exponential form, we extract the spin correlation length, which turns out 
to be weakly $T$-dependent and never exceeds a lattice spacing, in 
qualitative agreement with~\cite{[16]}. The resulting physical picture is 
that of a strongly fluctuating spin system, with extreme short-range AF 
correlations and {\it no} long-range order down to $T=0$.  This  
agrees with the absence of magnetic order found experimentally, as also 
with the fact that the CW constant corresponds to $S=1/2$, and that 
$\Theta_{CW}<0$.

A potentially interesting situation now occurs: the carrier hopping, as
mentioned above, reflects the $A_{1g}$ spin correlations as well as  
intrinsic geometric frustration effects from the FFL structure, preventing the 
tendency to charge- and/or orbital ordering (COO). This explicitly realizes 
the mechanism for suppression of COO proposed by Fulde {\it et al.}~\cite{[10]}; however, 
in our approach, it does not lead to formation of $S=1/2$ and $S=1$ rings and 
chains. The difference from the CMR manganites~\cite{[11]} also has a 
consistent explanation within this picture: the frustrated kinetic energy  
selfconsistently prevents the possibility for the $A_{1g}$ spins to 
align (via ``double exchange'') ferromagnetically, in contrast to what 
happens in manganites~\cite{[17]}. 

Next, consider the $E_{g}$-band electrons strongly coupled to the $A_{1g}$ 
spins via a local (ferromagnetic) Hund's rule coupling:

\bn
\nonumber
H_{el} &=& -t\sum_{<ij>,\sigma}(c_{ia\sigma}^{\dag} c_{jb\sigma}+h.c.) 
+ U\sum_{i,a}n_{ia\uparrow}n_{ia\downarrow} \\
&+& U_{ab}\sum_{i}n_{ia}n_{ib}
- J_{H}\sum_{i}{\bf S}_{i}.({\bf \sigma}_{ia}+{\bf \sigma}_{ib}) \;,
\en
where $a$ and $b$ refer to the doubly degenerate $E_{g}$ states in the $t_{2g}$
sector, defined on the FFL.  With the strong $J_{H}$, the ``double exchange'' 
projection transforms the problem to that of spinless fermions, but the 
hopping is modulated by the underlying, frustrated spin correlations in 
the $A_{1g}$ band.  The Hamiltonian is,

\be
H_{el}=-\sum_{<ij>,a,b}t_{ij}({\bf S})(c_{ia}^{\dag}c_{jb}+h.c) 
+ U_{ab}\sum_{i}n_{ia}n_{ib} \;,
\ee
where we set $U, J_{H}\rightarrow\infty$.  Relabelling $c_{a}=c_{\uparrow}$ 
and $c_{b}=c_{\downarrow}$, we are left with a Hubbard-like model with a 
non-trivial hopping term:

\be
H_{el}=-\sum_{<ij>,\sigma}t_{ij}^{\sigma}({\bf S})(c_{i\sigma}^{\dag}
c_{j\sigma}+h.c) + U_{ab}\sum_{i}n_{i\uparrow}n_{i\downarrow} \;.
\ee

We use the $d=\infty$ approximation to solve the Hubbard-like model 
on the FFL.  This interesting problem has non-trivial
solution(s), related to the one-electron dispersion 
on the FFL.  In particular, close to $n=1$, non-Fermi liquid
behavior is expected due to the strong scattering off the completely flat 
bands.  Fortunately, for $LiV_{2}O_{4}$, one deals with an almost
quarter-filled band, rendering the flat-band singularities irrelevant.
The above electronic model is now solved using the iterated perturbation 
theory (IPT) at slightly less than quarter-filling and at finite 
$T$~\cite{[18]}.

Fig.~\ref{fig2} shows the evolution of 
the one-electron local spectral function, $\rho(\omega)$.  As $U_{ab}$ 
increases, a sharp collective Kondo-like peak appears and gets narrower 
around $\mu$.  The appearance of upper and lower ``Hubbard'' bands (these 
appear as shoulder-like features in the range $-2\le\omega\le 2$ in Fig.2), 
reflecting suppression of charge fluctuations near quarter-filling, is 
also clear.  Formation of heavy quasiparticles is reflected 
in the quasiparticle renormalization constant, $Z(\mu)$, defined as
$Z(\mu)=\left[ 1-\frac{d\Sigma'(\omega)}{d\omega}|_{\omega=\mu} \right]^{-1}$.

\begin{figure}[h]
\epsfxsize=3.4in
\epsffile{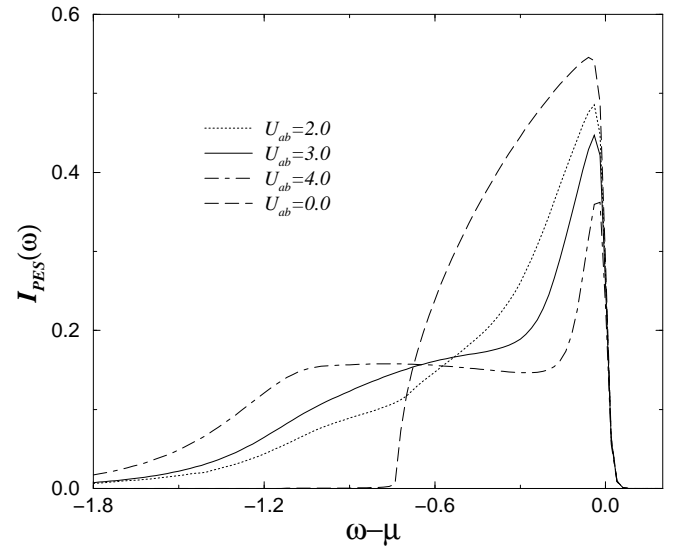}
\caption{ 
Integrated photoemission lineshapes 
of the Hubbard model on the fully-frustated pyrochlore lattice
for different values of $U_{ab}$.}
\label{fig3}
\end{figure}

$Z(\mu)$ decreases monotonically with increasing $U_{ab}$.  
That this correlated metallic state with heavy fermion mass 
($m^{*}/m=1/Z(\mu)$) is a FL is clear from the fact 
that Im$\Sigma(\omega \simeq \mu) = -b(\omega-\mu)^{2}$.  Importance 
of the filling is shown by the fact that charge fluctuations are 
enhanced, while spin fluctuations diminish in importance as $n$ is decreased 
further (not shown).  Our results are in complete agreement 
with those of Imai {\it et al.}~\cite{[19]} for a two-orbital 
Hubbard model with finite $U$ and $J_{H}$, and the HFFL behavior in our model
should persist when large, finite $U$ and $J_{H}$ are included.
The computed integrated 
photoemission lineshape in the HFFL phase is shown in Fig.~\ref{fig3}.  
Signatures of strong electronic correlations in the $E_{g}$ band are 
visible as shake-up features: a broad, incoherent, lower Hubbard band 
feature well separated from the narrow quasiparticle resonance.  This should 
provide evidence in favor of our modelling of the $E_{g}$ manifold.

  Finally, consider the question of selfconsistency due to the coupling of
the correlated carriers to the $A_{1g}$ spin fluctuations.  At 
low $T<T^{*}$, the dynamical spin susceptibility behaves like 
$\chi"(\Omega) \simeq (\Omega/T^{*})$ in the HFFL regime (see below).  
The leading-order correction to the self-energy of the correlated carriers 
is          
$\Sigma(i\omega_{n})=(J_{H}^{2}/\beta)\sum_{m}\chi(i\Omega_{m})
G(i\omega_{n}-i\Omega_{m})$.
In the HFFL regime, $G(\tau) \simeq (1/\tau)$ for long times, and 
$\Sigma(\omega)$ remains qualitatively unchanged from its FL-like form (see 
above) as long as 
$\chi"(\Omega) \simeq \Omega/T^{*}$.  This analysis ceases its validity 
above $T^{*}$, where the susceptibility gets increasingly dominated by 
fluctuating local moment contributions. Inserting the resulting 
$\Sigma(\omega)$ into the calculation for the susceptibility (notice that 
at low $T$, the spin susceptibility of the localized spins is gapped) leads 
again to the $\Omega/T^{*}$ form, apart from prefactors.  Thus, 
at low $T<T^{*}$, selfconsistency leads only
to a minor quantitative modification of our results. 

Let us discuss the implications of our calculation.  Since the magnetic 
correlation length is $T$-independent below $T^{*}<<J$ (corresponding to 
the point at which $\chi(T)$ shows a peak, see Fig.~\ref{fig1}) this sets 
the scale below which the hopping, $t$, can be treated as constant in 
$H_{b}$. Above $T^{*}$, the influence of the $A_{1g}$ spin correlations 
drive the system into a local-moment metallic regime, but below this scale, 
HFFL-like quasiparticles develop as shown above. We thus identify the low-$T$ 
crossover scale seen in experiments with the $T$-dependence of the frustrated 
spin dynamics in the $A_{1g}$ sector coupled to correlated $E_{g}$ carriers. 
This implies that the mechanism for HFFL behavior in our scenario is 
intimately linked to the local-moment magnetism of the FFL, and is
drastically different from the conventional view, where a band of 
uncorrelated~\cite{[20]} carriers collectively screens 
localized moments at every site.  Our calculation explicitly realizes the 
suggestions of Lacroix~\cite{[12b]}, but goes much further, showing 
clearly how short-ranged, spin-liquid-like local moment correlations in the 
$A_{1g}$ sector on the FFL are related to the onset of HFFL behavior.  In our
opinion, a mean-field-like decoupling {\it cannot} describe the crossover at
$T^{*}$ adequately, since the spin liquid behavior is driven precisely by 
strong fluctuations beyond the mean-field picture.

Given the HFFL metallic state below $T^{*}$, the electronic specific heat,
behaves like ${\cal C}_{el}(T) \simeq (T/T^{*})$, with a peak at $T^{*}$.  
The local 
dynamical spin susceptibility $J\chi"(\omega) \simeq (\omega/T^{*})$, so 
the NMR relaxation rate is Korringa like, $1/T_{1} \simeq (T/JT^{*})$.  
The uniform spin susceptibility $\chi \simeq O(1/J)$ at low $T$.  The 
calculation of the resistivity does not involve vertex corrections within 
DMFT.  At low $T$, in the HFFL phase, we obtain 
$\rho_{dc}(T) \simeq (T/T^{*})^{2}$, implying that the Kadowaki-Woods 
relation is obeyed, as in conventional heavy fermions. The Wilson ratio 
$W=T\chi/C_{el}\simeq O(T^{*}/J)$, again as in the traditional case.
All these results are indeed consistent with thermodynamic, transport and
magnetic measurements performed on $LiV_{2}O_{4}$.

A detailed calculation of the dynamical spin response of the FFL is a 
fascinating problem in itself, and is intimately linked to the details of 
the $T$-dependence of various quantities as $T$ crosses $T^{*}$. We plan to 
address this more detailed issue in a longer separate work. 

To conclude, a theoretical understanding of the various physical features of 
the frustrated, 3d, heavy fermion metal, $LiV_{2}O_{4}$ is proposed.
In particular, we have shown how the HFFL thermodynamics and
transport~\cite{[6]} can be reconciled with a
 magnetic response characteristic of
frustrated, insulating magnets~\cite{[7]} and how the experimentally observed 
crossover scale ($T^{*}$) is directly related to the $T$ dependence of the 
frustrated spin dynamics in the $A_{1g}$ sector.  To our knowledge, this is 
the first attempt that explicitly includes specific lattice structure, 
magnetic frustration and strong correlation effects within a single picture.  
Our picture is radically different from conventional ones, and is a concrete 
realization of the importance of geometrical frustration effects in driving 
HFFL behavior in $LiV_{2}O_{4}$.

MSL acknowledges the financial support of the SfB608 of the Deutsche 
Forschungsgemeinschaft.
LC and MSL wish to thank P. Fulde for his kind hospitality at the 
Max-Planck-Institute f\"ur Physik komplexer Systeme.

\end{document}